# Operationalizing Digital Self Determination

Dr. Stefaan G. Verhulst

The GovLab, Tandon School of Engineering, New York University, New York, USA

ISI Foundation, Turin, Italy

stefaan@thegovlab.org

October 25, 2022

**Abstract**

A proliferation of data-generating devices, sensors, and applications has led to unprecedented amounts of digital data. We live in an era of datafication, one in which life is increasingly quantified and transformed into intelligence for private or public benefit. When used responsibly, this offers new opportunities for public good. The potential of data is evident in the possibilities offered by open data and data collaboratives—both instances of how wider access to data can lead to positive and often dramatic social transformation. However, three key forms of asymmetry currently limit this potential, especially for already vulnerable and marginalized groups: data asymmetries, information asymmetries, and agency asymmetries. These asymmetries limit human potential, both in a practical and psychological sense, leading to feelings of disempowerment and eroding public trust in technology.

Existing methods to limit asymmetries (e.g., consent) as well as some alternatives under consideration (data ownership, collective ownership, personal information management systems) have limitations to adequately address the challenges at hand. A new principle and practice of digital self-determination (DSD) is therefore required.

The study and practice of DSD remains in its infancy. The characteristics we have outlined here are only exploratory, and much work remains to be done so as to better understand what works and what doesn't. We suggest the need for a new research framework or agenda to explore DSD and how it can address the asymmetries, imbalances, and inequalities—both in data and society more generally—that are emerging as key public policy challenges of our era.

# Operationalizing Digital Self Determination

Dr. Stefaan G. Verhulst[1]

**Introduction**

Our world is awash in data. Every day, some 2.5 quintillion bytes of data are generated,[2] and in 2020, approximately 64.2ZB of data was created or replicated.[3] According to the International Data Corporation (IDC), the amount of data created or replicated is growing at a compound annual growth rate of 23%, driven by the proliferation of Internet of Things (IoT) devices, remote sensors, and other data collection methods that are now deeply intertwined with virtually every aspect of people's professional and social lives.[4]

We have transitioned to a new *era of Datafication*—one in which human life is increasingly quantified, often without the knowledge of data subjects, and frequently transformed into intelligence that can be monetized for private or public benefit. At the same time, this new era offers tremendous opportunities: the responsible use and re-use of data can help address a host of apparently intractable societal and environmental problems, in large part by improving scientific research, public policymaking and decision-making.[5] Yet, it has also become increasingly clear that the datafication is simultaneously marked by a number of asymmetries, silos, and imbalances that are restricting the potential of data. This tussle—between potential and limits—is emerging as one of the central public policy challenges of our times.

In what follows, we outline a number of ways in which asymmetries are limiting the potential of datafication. In particular, we explore the notion of *agency asymmetry*, arguing that power imbalances in the data ecology are in effect disempowering key stakeholders, and as such undermining trust in how data is handled. We suggest that in order to address agency asymmetry (along with other forms of asymmetry), we *need a new principle and practice of digital self-determination*. This principle is built on the foundations of long-established philosophical, psychological, and legal principles of self-determination but updated for the digital era.

---

[1] The draft is based upon a presentation given to the October 2022 Residents of the Rockefeller Foundation Bellagio Center, and the author appreciates the input received from the residentes and the support of the Rockefeller Foundation to spent a month at the Center. The author is also grateful to Sampriti Saxena for her research assistance and the members of the International Network on Self Determination who provided comments to earlier drafts. https://idsd.network/)

[2] Branka Vuleta, "How Much Data Is Created Every Day? + 27 Staggering Stats," Seed Scientific (blog), October 28, 2021, https://seedscientific.com/how-much-data-is-created-every-day/.

[3] John Rydning, *Worldwide Global DataSphere and Global StorageSphere Structured and Unstructured Data Forecast, 2021–2025* (Needham: IDC Corporate USA, 2021), www.idc.com/research/viewtoc.jsp?containerId=US47998321.

[4] Ibid.

[5] Stefaan Verhulst, Andrew Young, Andrew J. Zahuranec, et al., *The Emergence of a Third Wave of Open Data: How To Accelerate the Re-Use of Data for Public Interest Purposes While Ensuring Data Rights and Community Flourishing* (Brooklyn: The GovLab, 2020), http://dx.doi.org/10.2139/ssrn.3937638.

In Part I, we explore how datafication has come about, and some of the asymmetries it has led to. Existing methods of addressing these asymmetries, we suggest, are insufficient; we need a new concept and practice of digital self-determination (DSD). We examine DSD in Part II, showing how it builds on a pre-existing intellectual tradition pertaining to self-determination. Part III contains a case study (on migration) that illustrates the notion of DSD, and Part IV contains some specific recommendations for operationalizing the practice of DSD.[6]

### I. Context: Digital Transformation and Datafication

The notion of datafication is sometimes conflated with big data. As we have elsewhere written, the two phenomena may be said to exist on a spectrum, but are in fact distinct.[7] In particular, datafication extends beyond a mere technical phenomenon and has what may be considered a sociological dimension. As Mejias and Couldry argue in the *Internet Policy Review*, datafication includes "the transformation of human life into data through processes of quantification," and this transformation, the authors further argue, has "major social consequences … [for] disciplines such as political economy, critical data studies, software studies, legal theory, and—more recently—decolonial theory."[8]

One aspect of datafication that is particularly relevant for our discussion here is that it is often intermingled with hierarchy and relationships of power. It exists, as Mejias and Couldry suggest, at "the intersection of power and knowledge."[9] This has tremendous implications for how data is accessed and distributed, in particular to the creation of data silos and asymmetries. We return to these challenges below. First, we consider the potential offered by datafication.

### A) Potential of Datafication: Reuse for Public Interest Purposes

---

[6] This paper originates from the deliberations and activities of the International Network on Digital Self-Determination. Learn more about the Network's work at idsd.network/.
[7] Stefaan G. Verhulst, "The Value of Data and Data Collaboratives for Good: A Roadmap for Philanthropies to Facilitate Systems Change Through Data," in *Data Science for Social Good*, ed. Massimo Lapucci and Ciro Cattuto (Cham: Springer, 2021), 9-27, https://doi.org/10.1007/978-3-030-78985-5_2.
[8] Ulises A. Mejias and Nick Couldry, "Datafication," *Internet Policy Review* 8, no. 4. (2019), https://policyreview.info/concepts/datafication.
[9] Ibid.

In order to understand the potential of datafication, we need to explore the possibilities offered by data reuse. Data reuse takes place when information gathered for one purpose is repurposed (often in an anonymized or aggregated) form for another purpose,, generally with an intended public benefit outcome. For example, location data collected by private telecommunication operators can be reused to understand human mobility patterns, which can help with public aid responses to mass migration events or during ecological crises. Likewise, clinical data held by medical practitioners can often be reused in the development of new drugs and treatments.

Such benefits are often realized through two key vehicles:

- **Open Data,** which involves data holders (typically in governmental and academic sectors) releasing data publicly, so that it can be "freely used, reused, and redistributed by anyone".[10]
- **Data collaboratives**, which are an emerging form of partnership, often between the public, academic and private sectors, that allow for data to be pooled and reused across data sets and sectors.[11] Data collaboratives can take a number of forms and allow data holders to provide access to data with other stakeholders for the public benefit without necessarily losing control or giving up a competitive advantage (this may be of particular concern to corporations and companies). In the example cited above, for instance, telecoms firms can provide access to their data with third-party researchers and responders while still maintaining their own stakeholder interests in the data.

### B) Key Challenge: Data Asymmetries

Open data and data collaboratives offer real potential to address some of the most intractable problems faced by society. When used responsibly and in the right data ecology, they can help policymakers by improving situational awareness, drawing clearer connections between cause and effect, enhancing predictive capabilities, and improving our understanding of the impact of critical decisions and policies.[12] All of this, when combined, can make a real and tangible difference in public decision-making. Similar benefits exist in the scientific community where greater access to data can lead to new research while allowing experiments to be reproduced and verified by anyone.[13]

---

[10] "What is Open Data?," *Open Data Handbook*, Open Knowledge Foundation, 2015, https://opendatahandbook.org/guide/en/what-is-open-data/.

[11] Andrew Young and Stefaan G. Verhulst, "Data Collaboratives," in *The Palgrave Encyclopedia of Interest Groups, Lobbying and Public Affairs*, eds. Phil Harris, Alberto Bitonti, Craig S. Fleisher and Anne Skorkjær Binderkrantz (Cham: Palgrave Macmillan, 2020), https://doi.org/10.1007/978-3-030-13895-0_92-1.

[12] Stefaan G. Verhulst, Andrew Young, Michelle Winowatan and Andrew J. Zahuranec, *Leveraging Private Data for Public Good: A Descriptive Analysis and Typology of Existing Practices* (Brooklyn: The GovLab, 2019), https://datacollaboratives.org/static/files/existing-practices-report.pdf.

At the moment, though, this potential is often held back. The key restriction stems from asymmetries in the way data is collected and, especially, stored (or hoarded). An era marked by unprecedented abundance—of data and other potential public goods—is also marked by vast disparities and hierarchies in how that abundance is distributed and accessed. Today, much of our data exists in silos, hidden from public view or usage, thus limiting the ability of policymakers, researchers, or other actors to benefit from its possibilities. In addition, the public is often left in the dark about how data is being collected, for what purpose, and how it is being used.

Three forms of asymmetry are worth highlighting:

- **Data asymmetries,** in which those who could benefit from access to data or draw out its potential are restricted from access;
- **Information asymmetries,** where there is a mismatch in awareness between data holders, data subjects, and potential users, meaning that data that could be useful is never sought or deployed; and
- **Agency asymmetries,** where data relationships between parties are marked by imbalances and hierarchies, meaning that one party—typically one that is already vulnerable and disenfranchised—is further disempowered. For instance, large quantities of data are collected on children daily, tracking their movements, communications, and more. Yet they (and their caregivers) have little to no agency over their data, and how it is used and later reused.[14]

The persistence of such asymmetries has a number of negative consequences. Most obviously, the potential public benefits of access to and reuse of data (e.g., through improved research or policymaking) are not fully realized. Lack of access to data may also contribute to bias in the analysis, especially if data hoarding leads to the exclusion of certain populations in the datasets.[15] In addition, the power imbalances mean that in effect an extractive relationship often exists between data subjects (e.g., citizens) and data holders (e.g., large companies), posing a number of practical and ethical consequences (observers have written of a sense of "data colonization").

---

[13] Ed Yong, "How Reliable Are Psychology Studies?," *The Atlantic*, August 27, 2015, https://www.theatlantic.com/science/archive/2015/08/psychology-studies-reliability-reproducability-nosek/402466/.
[14] Andrew Young and Stefaan G. Verhulst, "Why we need responsible data for children," The Conversation, March 23, 2020, https://theconversation.com/why-we-need-responsible-data-for-children-134052.
[15] This is especially of concern for developing countries and vulnerable groups, where a lack of access to representative data may further exclude these populations and amplify existing biases favoring specific groups or countries. See Seastedt et al., 2022 (https://doi.org/10.1371/journal.pdig.0000102).

All of this leads to a number of less obvious and more psychological, but no less insidious, consequences. The asymmetries and the sense of colonization lead to a feeling of disempowerment and a lack of autonomy, especially among populations that are already vulnerable, and this in turn erodes public trust in both technology and institutions—one of the defining problems of our times.

For all these reasons, it is essential that steps be taken to address the asymmetries that are at the heart of our data economy, in the process helping to unlock the value of the data age and spurring new forms of innovation in public decision-making. In Part II, below, we examine a *principle and practice of digital self-determination* that we believe is central to this process.

**C) Existing Methods of Rebalancing Asymmetries—and their Limitations**
First, though, we examine some existing methods either being deployed or considered to address these asymmetries. While these methods are well-intentioned and do sometimes have at least a marginal effect, we suggest that each has limitations and that, individually and collectively, they fail to address the underlying magnitude or scope of the problem.

   **i) Consent**

Today, the default approach to addressing information and power asymmetries involves the concept of informed consent. In this method, information about data handling policies is shared with data subjects, who then have the option of consenting whether or not to allow their data to be collected, accessed and (re)used. This has been the primary vehicle for providing data subjects with a "choice" since the widespread adoption of the Fair Information Practice Principles, approximately thirty years ago.[16] Yet, despite its widespread use and despite the fact that it provides the bedrock for many legislative efforts concerning data management,[17] informed consent has a number of shortcomings:

---

[16] U.S. Federal Trade Commission, *Privacy Online: A Report to Congress*, by Martha K. Landesberg, Toby Milgrom Levin, Caroline G. Curtin and Ori Lev (Washington, D.C.: FTC, 1998), https://www.ftc.gov/sites/default/files/documents/reports/privacy-online-report-congress/priv-23a.pdf.
[17] Including for example, the EU's GDPR and the OECD's Guidelines on the Protection of Privacy and Transborder Flows of Personal Data.

- **Binary:** Generally, opt-in or opt-out regimes dominate the practice of consent.[18] Yet such approaches tend to be binary–for or against collection or sharing—and thus inappropriately reductive. Some versions do allow for a greater level of granularity (i.e., more boxes to be checked), but even these fail to capture the true nuances and complexity of how data is collected, used, and reused.

- **Informational Shortcomings:** To truly confer agency, informed consent practices would need to convey a robust understanding of the nature, significance, implications, and risks of data collection, use and reuse.[19] For example, data subjects should be made aware of the immediate uses of their data, and also potential future uses. Such "rich information" is generally lacking, thus compromising citizens' ability to provide genuine consent.

- **Collective vs. Individual:** Informational shortcomings are exacerbated by the fact that consent policies are typically aimed at informing individuals about how their data will be used. In truth, however, data sets are often combined and repurposed in ways that have significant consequences for groups or communities. More responsible forms of consent would pay greater attention to the interests of communities.[20]

- **Limited Scope:** Finally, existing consent mechanisms are limited because much of the ethical and policy debate focuses on the scope of the original consent and whether reuse is permissible in light of that scope.[21] As a result, most consent regimes have a difficult time handling repurposing, which is so essential to fulfilling the potential of data. Recent years have witnessed the development of more open-ended consent models,[22] and several groups, such as the World Economic Forum,[23] have tried to improve on existing methods of consent to propose new methods. However, these too contain many ethical limitations, and can even act as bottlenecks to qualitative studies.[24]

---

[18] Yvonne de Man, et al., "Opt-in and opt-out consent procedures for the reuse of routinely recorded health data in scientific research and their consequences for consent rate and consent bias–A systematic review" (Preprint, 2022), https://doi.org/10.21203/rs.3.rs-1434893/v1.
[19] European Commission - Research Directorate-General, *Guidance for Applicants: Informed Consent* (Directive 2001/20/EC) (Brussels: European Commission, 2021), https://ec.europa.eu/research/participants/data/ref/fp7/89807/informed-consent_en.pdf.
[20] Leslie P. Francis and John G. Francis, "Data Re-Use and the Problem of Group Identity," University of Utah College of Law Research Paper No. 311, *Studies in Law, Politics and Society* 73 (2017), http://dx.doi.org/10.2139/ssrn.3371134.
[21] Ibid.
[22] Ibid.
[23] Kimberly Bella, Christophe Carugati, Cathay Mulligan and Marta Piekarska-Geater, *Data for Common Purpose: Leveraging Consent to Build Trust* (Cologny: World Economic Forum, 2021), https://www3.weforum.org/docs/WEF_Data_for_Common_Purpose_Leveraging_Consent_to_Build_Trust_2021.pdf.

### ii) Alternative Consent Mechanisms—and their Shortcomings

In part due to these shortcomings, some have suggested "post-consent privacy," while others have suggested the establishment of alternative rights and technologies. However, each of these also contains certain limitations.[25]

- **Data Ownership Rights:** One approach is to treat data as the private property of data subjects. While in theory this could enhance agency, it poses the serious problem of undermining the public good properties of data. Data is non-rivalrous, non-excludable, and non-depletable, making it by definition a public good.[26] While ownership of data may appear to solve problems related to consent and control, it raises serious concerns regarding the marketization and commodification of data.[27] In truth, a lack of clarity regarding the notion of *ownership* when it comes to data means that it cannot be treated as solely a public or private good.

- **Collective Ownership:** Ownership of data can be at the level of the individual, the community, or group. Group-level ownership has most commonly been explored under the rubric of "data sovereignty," which places data under the jurisdictional control of a single political entity.[28] Collective ownership poses many of the same challenges as those posed by individual ownership, notably those associated with the privatization of a public good. In addition, a lack of operationalization in terms of clear and enforceable legal frameworks around data ownership makes it difficult to establish or operationalize data sovereignty.[29] National or subnational jurisdictions are often in conflict, and different areas of law operate differently. For example, while intellectual property rights[30] protect certain aspects of data reuse, criminal law[31] may interpret reuse as a form of theft. Overall, the absence of a legal framework allows data subjects to be exploited, while also limiting the effective reuse of data for the public good.

---

[24] Sara Mannheimer, "Data Curation Implications of Qualitative Data Reuse and Big Social Research," *Journal of eScience Librarianship* 10, no. 4 (2021), https://doi.org/10.7191/jeslib.2021.1218.

[25] Solon Barocas and Helen Nissembaum, "Big Data's End Run around Anonymity and Consent," in *Privacy, Big Data, and the Public Good: Frameworks for Engagement,* eds. Julia Lane, Victoria Stodden, Stefan Bender and Helen Nissenbaum (Cambridge: Cambridge University Press, 2014), https://doi.org/10.1017/CBO9781107590205.004.

[26] Patrik Hummel, Matthias Braun and Peter Dabrock, "Own Data? Ethical Reflections on Data Ownership," *Philosophy & Technology* 34 (2021), https://doi.org/10.1007/s13347-020-00404-9.

[27] Jonathan Montgomery, "Data Sharing and the Idea of Ownership," *The New Bioethics* 23, no. 1 (2017), https://doi.org/10.1080/20502877.2017.1314893.

[28] Théodore Christakis, *"European Digital Sovereignty": Successfully Navigating Between the "Brussels Effect" and Europe's Quest for Strategic Autonomy* (Grenoble: CESICE, 2020), http://dx.doi.org/10.2139/ssrn.3748098.

[29] Patrik Hummel, Matthias Braun and Peter Dabrock, "Own Data? Ethical Reflections on Data Ownership," *Philosophy & Technology* 34 (2021), https://doi.org/10.1007/s13347-020-00404-9.

- **Personal Information Management Systems:** Personal Information Systems (PIMS)[32] are sometimes proposed as alternative systems of data management to empower individuals with greater control over their personal data. PIMS are typically centralized or decentralized systems through which individuals can choose to share (or not share) their personal data. This method also faces some important limitations.

   First, there is a danger that, rather than conferring control on individual subjects, PIMS will simply transfer control to owners and operators of large PIMS systems. In this argument, a PIMS-based system of consent will end by replicating (and perhaps aggravating) existing hierarchies and asymmetries.

   Second, PIMS remain highly susceptible to the many vulnerabilities of the existing data ecosystem. In particular, they are prone to hacking and breaches, and are only as robust as the surrounding legal and policy ecosystem that protects how data is collected, stored, and shared.[33]

   Finally, the adoption of PIMS has been stunted by a lack of adequate use cases and, consequently, an insufficiently proven business case.[34] Without stronger models and stress-tested best practices, the potential of PIMS remains more conceptual than proven.

## II. Need for new principle: Digital Self Determination

---

[30] Sara Mannheimer, "Data Curation Implications of Qualitative Data Reuse and Big Social Research," *Journal of eScience Librarianship* 10, no. 4 (2021), https://doi.org/10.7191/jeslib.2021.1218.
[31] Kathleen Liddell, David A. Simon and Anneke Lucassen, "Patient data ownership: who owns your health?," *Journal of Law and the Biosciences* 8, no. 2 (2021), https://doi.org/10.1093/jlb/lsab023.
[32] "Personal Information Management System," *European Data Protection Supervisor,* European Union, 2021, https://edps.europa.eu/data-protection/our-work/subjects/personal-information-management-system_en.
[33] "Personal information management systems: A new era for individual privacy?," Privacy Tech, International Association of Privacy Professionals (IAPP), March 21, 2019, https://iapp.org/news/a/personal-information-management-systems-a-new-era-for-individual-privacy/.
[34] Heleen Janssen and Jatinder Singh, "Personal Information Management Systems," *Internet Policy Review* 11, no. 2 (2022), https://policyreview.info/glossary/personal-information-management-systems.

All these shortcomings, of both existing and hypothetical methods of agency, call out for a new approach to addressing the asymmetries of our era. Our proposed solution rests on the principle of digital self-determination (DSD). As noted, DSD is built on the foundations of existing practices and principles about self-determination. As a working definition, we propose the following:

> *Digital Self-Determination is defined as the principle of respecting, embedding, and enforcing people's and people's agency, rights, interests, preferences, and expectations throughout the digital data life cycle in a mutually beneficial manner for all parties involved.*

### A) The Concept of Self Determination

The above definition may be a relatively new concept, but it stems from a historical body of exploration that involves philosophy, psychology, and human rights jurisprudence. The term "self-determination" is often attributed to the German philosopher Immanuel Kant, who wrote in the 19th century about the importance of seeing humans as "moral agents" who would respect rules over their own needs and emotions because of an innate feeling of social "duty."[35] Regardless of personal feelings, he argued, humans "have the duty to respect dignity and autonomy"—the *self-determination*—of others.[36] More generally, Kant's philosophy affirmed the importance of treating individuals as ends rather than means, and of the importance for individuals to be able to remain *eigengesetzlich* (autonomous).[37]

Self-determination can also be explored through the prism of psychology, where the ability to make decisions for oneself is often considered central to people's motivations, well-being, and fulfillment.[38] For instance, Ryan and Deci (1980) explore self-determination theory, which argues that there is a dichotomy between "automated," instinctive behaviors and consciously "self-determined behaviors" to achieve a specific outcome.[39]

---

[35] Nydia Remolina and Mark Findlay, "The Paths to Digital Self-Determination–A Foundational Theoretical Framework," *Singapore Management University Centre for AI & Data Governance Research Paper* 03/2021 (2021), https://dx.doi.org/10.2139/ssrn.3831726.

[36] Immanuel Kant, "The Metaphysics of Morals," in *Practical Philosophy*, ed. Mary J. Gregor (Cambridge: Cambridge University Press, 1997), https://doi.org/10.1017/CBO9780511813306.

[37] Kimberly Hutchings, "The question of self-determination and its implications for normative international theory," *Critical Review of International Social and Political Philosophy* 3, no. 1 (2000): 91-120, https://doi.org/10.1080/13698230008403304.

[38] Richard M. Ryan and Edward L. Deci, "Self-Regulation and the Problem of Human Autonomy: Does Psychology Need Choice, Self-Determination, and Will?," *Journal of Personality* 74, no. 6 (2006): 1557-1586, https://doi.org/10.1111/j.1467-6494.2006.00420.x.

[39] Edward L. Deci and Richard M. Ryan, "Self-determination Theory: When Mind Mediates Behavior," *The Journal of Mind and Behavior* 1, no. 1 (1980): 33-43, https://www.jstor.org/stable/43852807.

Finally, it is also worth considering international law, which upholds the notion of self-determination as it applies to both states and their constitutive members, i.e., individuals. Self-determination is for instance closely associated with the decolonization movement, as well as with movements for the autonomy and independence of indigenous people. In 2007, the UN Declaration on the Rights of Indigenous Peoples (UNDRIP) acknowledged the right of peoples to practice customs and cultures "without outside interference" while also taking "part in the conduct of public affairs at any level,"[40] thereby asserting the fundamental importance of autonomy for individuals and groups of people. The basis for this importance can be extended even further back, to Article 1 of the 1966 International Covenant on Economic, Social and Cultural Rights and the International Covenant on Civil and Political Rights, which state that: "All peoples have the right to self-determination. By virtue of that right they freely determine their political status and freely pursue their economic, social, and cultural development."[41]

### B) Digital Self Determination

The increased digitization and datafication of society lead us to extend these notions of self-determination to a concept of digital self-determination. DSD includes the following key aspects and components:

*DSD is mainly concerned with agency about data*
First, with the advent of digital technologies rapidly advancing the collection, storage and use of data, the need for DSD has become increasingly more pressing in recent years. As a concept, it rests essentially on the understanding that, in a digital society, data and individuals are not separate entities, but mutually constitutive.[42] Thus control over one's data representation is fundamentally a matter of individual agency and liberty.

*DSD has both an individual and collective dimension*
Second, like physical self-determination, DSD has an external *collective dimension* that accounts for the influence that other people, peoples and communities have on the virtual social self. Vice versa, DSD also has an internal *individual dimension* that defines the whole online self as the sum of three elements: one's virtual persona, one's data, and data about oneself. Each of these elements is essential in considering the notion of DSD and how best to apply it.

---

[40] Daniel Thürer and Thomas Burri, "Self-Determination," in *Max Planck Encyclopedias of International Law*, ed. Rüdiger Wolfrum (Heidelberg: Max Planck Institute for Comparative Public Law and International Law, 2008), https://opil.ouplaw.com/view/10.1093/law:epil/9780199231690/law-9780199231690-e873?prd=EPIL.
[41] UN General Assembly, *International Covenant on Economic, Social and Cultural Rights (Resolution 2200A - XXI)*, December 16, 1966, www.ohchr.org/en/instruments-mechanisms/instruments/international-covenant-economic-social-and-cultural-rights.
[42] See Remolina and Findlay (2021), who propose a five-step DSD framework.

*DSD can especially benefit the vulnerable, marginalized, and disenfranchised*
Third, DSD is ethically desirable for the way in which it protects subject rights, whether individually or collectively. It is also worth noting that the need for DSD is particularly important to protect the rights of society's most marginalized and disenfranchised—those who are typically less included in, and aware of, the emerging processes of social datafication. These populations are often the most vulnerable to digital asymmetries, and already excluded from various aspects of social and economic life in the digital era. As such, there is a strong redistributive dimension to DSD.

*DSD can leverage existing practices of principled negotiation*
Fourth, the notion of determination creates a new avenue for negotiation beyond traditional institutional levers. Based on learnings from principled negotiation theory, DSD can help to establish objective criteria, focus on specific individual and collective interests, and unite common options.[43] Negotiations framed around objective criteria are more efficient and productive, as they highlight mutual gain and help balance power asymmetries. DSD plays a role in establishing objective criteria by empowering individuals to advocate for and establish their pre-existing interests in a broader negotiation process.[44] This not only helps focus on important interests that may otherwise be ignored, but also helps to frame negotiations in a way that unites common interests to achieve more fair outcomes.

*DSD will need to be flexible and context-specific, yet enforceable*
Finally, it is important to emphasize that DSD cannot be achieved in a strictly pro-forma or prescriptive way but will often need to be approached in a voluntary, contextual and participatory manner (in this sense, it is conceptually reminiscent of notions of self-regulation). Such an approach of "productive ambiguity" aligns more closely to the principles of self-determination and can help to ensure the successful adoption of DSD in the long run. How DSD is implemented will depend on what data is being handled, the stage of the data lifecycle that is being considered, who the actors are, and how interests are being addressed. Each context will call for its own set of stakeholders, processes, and systems. At the same time, because of the well-established weaknesses of enforcement in self-regulatory contexts, special attention will need to be given to how to enforce the negotiated conditions of DSD.

---

[43] Tanya Alfredson and Azeta Cungu, *Negotiation Theory and Practice: A Review of the Literature* (Rome: FAO, 2008), https://www.fao.org/3/bq863e/bq863e.pdf.
[44] Ibid.

### III. Case-Study: Migrants

The concept of DSD can be productively explored through case studies. In this section, we explore an example related to migrant populations, the challenges they face concerning data, and how DSD can protect and help them flourish.[45]

Migrant populations today account for an estimated 3.6% of the world's population, a number that continues to grow as global crises increase.[46] Already in 2022, the COVID-19 pandemic, the war in Ukraine, and the floods in Pakistan have displaced millions of people. As the number of migrants around the world increases, so too do the number of technologies associated with their journeys. These tools generate and use huge amounts of data, often without the express consent of the data subjects.[47] Consider the following examples:

- In 2013, at a refugee camp in Malawi, the UNHCR launched the Biometric Identity Management System (BIMS). This system holds "body-based" identifiers—including fingerprints, iris scans, and facial scans—to accredit refugees and grant a service access to food rations, housing, and spending allowances.[48] Additionally, UNHCR employs blockchain to link individuals with transaction data.

- The EUMigraTool uses data from migrants sourced from video content, web news, and social media text content to generate modeling and forecasting tools to help manage migrants' arrival and support needs in a new country.[49] Through its algorithms, the tool can help predict migration flows and detect risks and tensions related to migration, allowing migration service organizations to prepare for the appropriate amount of human and material resources needed when responding to a migration event.

---

[45] Based upon the findings of a studio we conducted in 2021/2022 with the International Network on Digital Self Determination and the Big Data for Migration Alliance. To learn please see this article.
[46] *World Migration Report 2020*, eds. Marie McAuliffe and Binod Khadria (Geneva: International Organization for Migration, 2020), https://publications.iom.int/system/files/pdf/wmr_2020.pdf.
[47] Kenneth Neil Cukier and Victor Mayer-Schoenberger, "The Rise of Big Data: How It's Changing the Way We Think About the World," *Foreign Affairs*, May/June 2013, https://www.foreignaffairs.com/articles/2013-04-03/rise-big-data ; Jos Berens et al., "The Humanitarian Data Ecosystem: the Case for Collective Responsibility," *Stanford Center on Philanthropy and Civil Society (PACS)* (2016), https://pacscenter.stanford.edu/publication/the-humanitarian-data-ecosystem-the-case-for-collective-responsibility/.
[48] *Biometric Identity Management System: Enhancing Registration and Data Management* (Geneva: UN Refugee Agency (UNHCR), 2015, https://www.unhcr.org/550c304c9.pdf.
[49] "EUMigraTool," *IT Flows*, 2022, https://www.itflows.eu/eumigratool/.

- X2AI, a mental healthcare app, developed 'Karim,' a Chatbot to provide virtual psychotherapy[50] to Syrians in the Zaatari refugee camp. The non-profit Refunite[51] assists refugees in locating missing family members via mobile phone or computer, and currently has over 1 million registered users. And Mazzoli et al. (2020)[52] demonstrate how geolocated Twitter data can help identify specific routes taken, as well as areas of resettlement, by migrants during migrant crises.

These are just a few examples that illustrate how data is both generated by migrant movements and also used to channel aid, resettle populations, and generally inform the policy response. Without a doubt, there are many potential benefits to such usage. Much of the generated data can be leveraged in the pursuit of evidence-based policy-making to alleviate the sufferings and marginalization of this vulnerable population.

But as in virtually every other aspect of our digital era, data also poses a threat to migrant populations, notably by potentially infringing upon their rights and creating new power structures and inequalities.[53] Migrants face power imbalances when it comes to agency over their data, choice in how their data is used, and control over who has access to their data.[54] These asymmetries are often further exacerbated by a lack of digital literacy, limiting migrants' ability to use digital tools to achieve self-determination. For example, with little to no agency over the use of their data, migrant populations are often exploited as test subjects for new technologies, rather than benefiting from these rapid developments.[55] In addition, in many use cases, there exists a very weak framework for how data is collected, stored, and generally used, leading to ample scope for abuses.

---

[50] Nick Romeo, "The Chatbot Will See You Now," Annals of Technology, *The New Yorkers,* December 25, 2016, https://www.newyorker.com/tech/annals-of-technology/the-chatbot-will-see-you-now.
[51] "Refunite." *Refunite*, https://refunite.org/.
[52] Mattia Mazzoli et al., "Migrant mobility data flows characterized with digital data," *PLoS ONE* 15, no. 3 (2020), https://doi.org/10.1371/journal.pone.0230264.
[53] Jessica Bither and Astrid Ziebarth, AI, Digital Identities, Biometrics, Blockchain: A Primer on the Use of Technology in Migration Management (Washington, D.C.: The German Marshall Fund, 2020), https://www.gmfus.org/news/ai-digital-identities-biometrics-blockchain-primer-use-technology-migration-management ; *The use of digitalisation and artificial intelligence in migration management: Joint EMN-OECD Inform* (Brussels: European Migration Network, 2022), https://www.oecd.org/migration/mig/EMN-OECD-INFORM-FEB-2022-The-use-of-Digitalisation-and-AI-in-Migration-Management.pdf.
[54] Stefaan Verhulst, Marine Ragnet and Uma Kalkar, "Digital Self-Determination as a Tool for Migrant Empowerment," *Big Data for Migration* (blog), May 26, 2022, https://data4migration.org/articles/digital-self-determination-studio-digital-self-determination-as-a-tool-for-migrant-empowerment/index.html.
[55] Petra Molnar, "New technologies in migration: human rights impacts," the ETHICS issue, *Forced Migration Review,* June 2019, https://www.fmreview.org/ethics/molnar ; Aaron Martin et al., "Digitisation and Sovereignty in Humanitarian Space: Technologies, Territories and Tensions," *Geopolitics* (2022), https://doi.org/10.1080/14650045.2022.2047468.

DSD may offer some potential solutions to these growing asymmetries. Applied responsibly, DSD can help address power and agency asymmetries between migrants and various stakeholders by empowering migrants with the ability to control how their data is collected, stored and used. It also creates avenues for negotiation, whereby trusted intermediaries can advocate for migrants and for other stakeholders in shared ecosystems. DSD's focus on the "self" helps direct discussions and frameworks around the unique experience of various migrant populations, thus making DSD more effective in addressing the specific vulnerabilities and contextual factors facing different populations today. DSD is also useful because it can help engage migrants in the process of data generation, collection, use and reuse, thus opening avenues for their engagement in the policy process and widening the range of insights brought to bear on the data policy process.[56]

These are just some of the ways in which DSD can be useful in addressing a pressing global socio-economic problem. In the next section, we examine how these insights can be operationalized more generally, across sectors and domains.

## IV. Operationalizing Digital Self-Determination

In order for DSD to have a social impact and help mitigate the asymmetries of our era, it is critical for theory to be translated into practical implementation. This represents a critical step in moving from concept to concrete policy implementation. As always within the data ecology, the task is not simply to blindly apply the concepts explored above but to understand how to do so *responsibly*—in a manner that maximizes agency, and balances the potential benefits with the possible harms of any possible policy or technical intervention.

Responsible implementation of DSD can be explored through a four-pronged framework: processes, people and organizations, policies, and products and technologies.

---

[56] Hannah Chafetz, Uma Kalkar, Marine Ragnet, Stefaan Verhulst and Andrew J. Zahuranec, "How Can We Ensure the Digital Self-Determination of Migrants," *Big Data for Migration (blog),* July 18, 2022, https://data4migration.org/articles/digital-self-determination-studio-how-can-we-ensure-the-digital-self-determination-of-migrants/index.html.

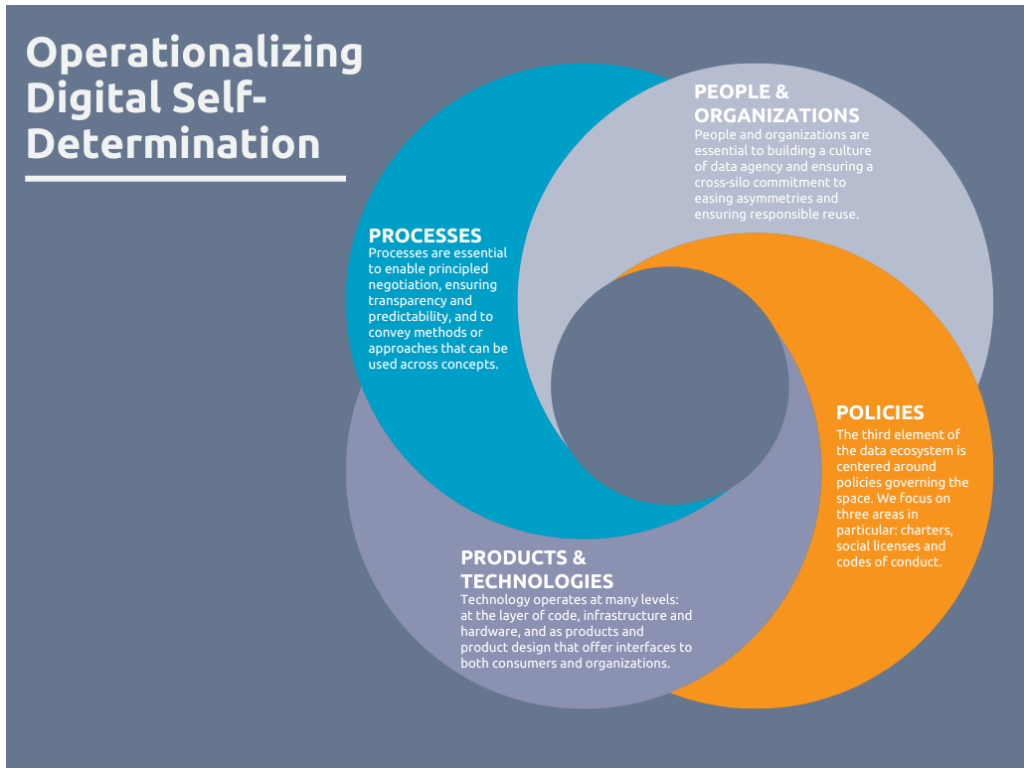

*Operationalizing Digital Self-Determination: a four-pronged framework*

### A) Processes: Exploring the role of Data Assemblies

Processes are essential to enable principled negotiation, ensuring transparency and predictability, and conveying methods or approaches that can be used across concepts. Some key processes to can be considered in the operationalization of DSD include citizen data commons,[57] citizen engagement programs,[58] public deliberations, and participatory impact assessments.[59]

One process holding particular potential involves the use of data assemblies, or citizen assemblies or juries around the reuse of data. Data assemblies bring together policymakers, data practitioners, and key members of communities to co-design the conditions under which data can be reused, as well as various other associated issues.[60] As an example, in 2020, the GovLab launched The Data Assembly initiative, a citizens assembly based in New York City. Through this approach, participants were able to understand how different stakeholders perceive the challenges and risks of data reuse, as well as the diverse value propositions data reuse promises each actor. Among the key lessons of this project was the finding that data assemblies not only create space for public engagement but also offer avenues through which data practitioners can secure responsibly informed consent from the public—an essential step in building a more trusted and engaged data ecology.[61]

---

[57] Liton Kamruzzaman, "Net zero precincts: Citizen data commons and technological sovereignty," *Monash University*, https://www.monash.edu/mada/research/net-zero-precincts.
[58] "Citizen Engagement and Innovative Data Use for Africa's Development (DataCipation)," GIZ, 2021, https://www.giz.de/en/worldwide/98533.html.
[59] Reema Patel et al., *Participatory data stewardship: A framework for involving people in the use of data* (London: The Ada Lovelace Institute, 2021), https://www.adalovelaceinstitute.org/wp-content/uploads/2021/11/ADA_Participatory-Data-Stewardship.pdf.
[60] *The Data Assembly*, The GovLab, 2020, https://thedataassembly.org/.

### B) People and Organizations

People and organizations also play a key role in operationalizing DSD. Among other functions, they are essential to building a culture of data agency and ensuring a cross-silo commitment to easing asymmetries and ensuring responsible reuse. People and organizations are in essence the building blocks of responsible data use and reuse.

To operationalize DSD, two critical functions or roles for individuals and groups need to be highlighted:

### i) Data Stewards

Individuals or groups of individuals occupying the emerging function of data stewards within organizations play important roles in facilitating DSD and responsible data reuse. A data steward is a leader or team "empowered to create public value by re-using their organization's data (and data expertise); identifying opportunities for productive cross-sector collaboration and responding proactively to external requests for functional access to data, insights or expertise."[62] Their roles and responsibilities include engaging with and nurturing collaborations with internal and external stakeholders, promoting responsible practices, implementing governance processes, and communicating insights with broader audiences.[63] Data stewards are key actors in enabling the operationalization of DSD due to their ability to promote the adoption of processes and practices that empower data subjects to effectively assert agency.

### ii) Data Intermediaries

---

[61] Andrew Zahuranec, Andrew Young and Stefaan G. Verhulst, "How can stakeholder engagement and mini-publics better inform the use of data for pandemic response?," *Participo* (blog), February 19, 2021, https://link.medium.com/Y9COUWwQnub.

[62] Stefaan G. Verhulst, Andrew J. Zahuranec, Andrew Young and Michelle Winowatan, *(Re-)Defining the Roles and Responsibilities of Data Stewards for an Age of Data Collaboration* (Brooklyn: The Governance Lab, 2020), http://www.thegovlab.org/static/files/publications/wanted-data-stewards.pdf.

[63] Stefaan G. Verhulst, "Data Stewardship Re-Imagined – Capacities and Competencies," *Data Stewards Network* (blog), October 8, 2021, https://link.medium.com/dJWWpcvQnub.

If data stewards facilitate responsible data reuse, then data intermediaries are emerging as potential solutions to the challenges posed by unbalanced collective bargaining. These individuals or teams mediate transactions between the supply and demand of data. For example, they may help match a private sector organization that currently stores large sets of (siloed) data with a non-profit organization that can apply that data toward the public good. In the context of DSD, data intermediaries can help balance the need for data subjects to maintain agency over their own data while at the same time enabling a robust data-sharing ecosystem. Data intermediaries do pose certain challenges, notably the risks of higher transaction costs and the creation of new power asymmetries. These can be mitigated by regulatory frameworks that ensure that data intermediaries remain neutral, fair, and secure.[64]

### C) Policies

The third element needed to operationalize DSD relates to governance and policies. We focus on three areas in particular: charters, social licenses, and codes of conduct.

### i) Charter

DSD is, in essence, about balancing power and agency asymmetries. These imbalances center around agency, choice, and participation. The creation of a charter (or statements of intent) nuanced by the types of actors, data collection and data usage may provide a potential policy-based solution to this imbalance. Past data charters, such as the International Open Data Charter[65] or the Inclusive Data Charter,[66] could act as models for a DSD charter. Such a charter would serve as a unifying step to define, scope, and establish DSD for data actors. In order to be most effective, it is important that any charter for DSD takes a life cycle, multi-stakeholder approach. Additionally, a charter ought to be context specific and human-centered to ensure that the rights of the data subject are protected.

### ii) Social License

Social licenses are another policy tool that can help operationalize DSD. A social license, or social license to operate, captures multiple stakeholders' acceptance of standard practices and procedures, across sectors and industries.[67] Social licenses help facilitate responsible data reuse by establishing standards of practice for the sector as a whole. They also empower individual actors to exert more proactive control over their data, which is critical to the adoption of DSD.

---

[64] "Governing Data Intermediaries – The Data Governance Act: Principles, Frictions, and Perspectives," *Florence School of Regulation (EUI)*, April 13, 2022, https://fsr.eui.eu/governing-data-intermediaries-the-data-governance-act-principles-frictions-and-perspectives/.
[65] "The International Open Data Charter," *Open Data Charter,* https://opendatacharter.net/.
[66] "Inclusive Data Charter," *Global Partnership for Sustainable Development Data*, 2018, https://www.data4sdgs.org/initiatives/inclusive-data-charter.
[67] Will Kenton, "Social License to Operate (SLO)," *Investopedia*, May 31, 2021, https://www.investopedia.com/terms/s/social-license-slo.asp.

Broadly, there exist three approaches to secure social licenses for data reuse: public engagement; data stewardship; and regulatory frameworks.[68] In the context of DSD, data stewards play an especially important role, given their position as facilitators of responsible data reuse.

### iii) Codes of Conduct

Codes of conduct are another policy element that can help operationalize DSD. A code of conduct interprets a policy and lays the foundation for its implementation in a specific sector.[69] By bringing together diverse stakeholders, or "code owners," a code of conduct is able to account for the many different interests as well as technical and logistical requirements at play in the ecosystem. Moreover, a code of conduct can lead to the creation of a monitoring body, which is responsible for ensuring compliance, reviewing and adapting procedures, and sanctioning members who break the code. The monitoring body not only helps implement the code in a dynamic and effective manner, but can also foster secure data sharing by acting as a third-party intermediary.

### D) Products and Technological Tools

While the preceding elements are largely focused on human or human-initiated processes, it is important to recognize that technology also plays an important role in operationalizing DSD. Technology operates at many levels: at the layer of code, infrastructure and hardware, and as products and product design that offer interfaces to both consumers and organizations. User-led design experience,[70] informed by the needs of both consumers and larger beneficiaries of data, can help implement DSD principles in practice by promoting digital access and action across stakeholders.[71]

One technological product that can play a significant role is a trusted data space. A data space can be defined as an "organizational structure with technical and physical components that connects data users and data providers with sources of data."[72] A trusted data space gives stakeholders a degree of control over this space and thus over their data, while still encouraging sharing practices. This balance between agency and the right to reuse is a step toward DSD, as it aims to protect and empower data subjects without hampering open data.

---

[68] Stefaan G. Verhulst and Sampriti Saxena, "The Need for New Methods to Establish the Social License for Data Reuse," *Data & Policy Blog* (blog), May 20, 2022, https://link.medium.com/uZ9layFQnub.

[69] Mathias Vermeulen, "The Keys to the Kingdom," *Knight First Amendment Institute at Columbia University Law and Political Economy Project essay series* (2021), https://knightcolumbia.org/content/the-keys-to-the-kingdom.

[70] Reema Patel et al., *Participatory data stewardship: A framework for involving people in the use of data* (London: The Ada Lovelace Institute, 2021), https://www.adalovelaceinstitute.org/wp-content/uploads/2021/11/ADA_Participatory-Data-Stewardship.pdf.

[71] Sandra Ponzanesi, "Migration and Mobility in a Digital Age: (Re)Mapping Connectivity and Belonging," *Television & New Media* 20, no. 6 (2019), https://doi.org/10.1177/1527476419857687.

**V. Considerations and Reflections**

In addition to outlining these four areas of operationalization, we wish to offer some additional considerations and observations on the operationalization of DSD—both in its current incipient state, and in the more fleshed-out version that may yet emerge.

**A) Life Cycle Approach:**

In order for that more fleshed-out, operational version to take shape, it is going to be essential to identify opportunities for DSD that exist at each stage of the data life cycle. The data life cycle follows data from its creation to its transformation into an asset across five stages: collection, processing, sharing, analyzing, and using.[73] Each stage of the process could benefit from DSD. As we've seen above, the collection and the sharing stages of the life cycle must overcome challenges posed by agency asymmetries, which can be mitigated by the principle and practice of DSD. Similarly, during the processing, analysis, and use of data, opportunities for DSD emerge in terms of the use (and reuse) of data and the sharing of insights.

By taking a data life cycle approach to DSD, stakeholders will also be well-positioned to account for the variety of asymmetries, both in terms of data and in terms of power, that exist across different levels of the ecosystem. Starting from the level of the individual and extending all the way to the national stage, each actor defines different notions of self-determination to address unique asymmetries. The life cycle approach allows for stakeholders to reflect on and respond to the varied asymmetries that exist at each stage to help achieve a greater balance in power.

**B) Symmetric Relationships**

---

[72] Federal Department of the Environment, Transport, Energy and Communications (DETEC) and the Federal Department of Foreign Affairs (FDFA), *Creating trustworthy data spaces based on digital self-determination: Report from the DETEC and the FDFA to the Federal Council on 30 March 2022* (Bern: The Swiss Confederation, 2022): 16, https://digitale-selbstbestimmung.swiss/home/en/245-2/.

[73] Andrew Young, Andrew Zahuranec and Stefaan Verhulst, "A Layered Approach to Documenting How the Third Wave of Open Data Can Provide Societal Value," *Open Data Policy Lab* (blog), August 21, 2021, https://opendatapolicylab.org/articles/the-onion-model-a-layered-approach-to-documenting-how-the-third-wave-of-open-data-can-provide-societal-value/index.html.

As we have seen, DSD offers many benefits. One of the most important is its role in building symmetric relationships between stakeholders by re-balancing existing power and agency asymmetries. Symmetric relationships are important ethically, but they also have the potential for greater stability in the long term, and help prevent the exploitation of weaker parties by stronger ones.[74]

In the context of a data ecosystem, symmetric relationships can help data subjects more effectively leverage their self-determination to exert a greater degree of control over the ways in which their data is used and reused. This is especially important to vulnerable minorities, who may also be disempowered in other ways. Future systems ought to be designed with these permanent minorities in mind within the broader context of human rights, justice and democracy.[75]

### C) DSD and Disintermediation

Finally, DSD can help limit the creation of new power asymmetries by preventing the emergence of new chokepoints and loci of control in the form of new intermediaries. By returning power and agency to individual stakeholders, the need for dominant intermediaries is minimized. One important consequence is a lowering in the risk of new power imbalances, which so often stem from the disproportionate power of intermediaries.

Removing dominant intermediaries also simplifies the process of developing symmetric relationships, which are easier to achieve without the role of middlemen. In this way, disintermediation can play a vital role in increasing subject agency, and in empowering data subjects to exert control over their own data while also promoting safe data sharing.

## VI. Further Research and Action

Self Determination is a historical concept with an impressive intellectual and juridical pedigree. It is imperative that this concept, like so many others, be updated to the digital era. The notion of DSD we have outlined here is preliminary, more exploratory and conjectural than finalized. It sets the foundations for a fuller exploration of DSD, as well as the vital roles of agency, asymmetries, and other power dynamics within the data ecology. Our hope is that this paper sets an agenda or framework for further action and research into, and ultimately operationalization of, DSD across sectors and industries in our era of rapid and unrelenting datafication.

---

[74] Frank R. Pfetsch, "Power in International Negotiations: Symmetry and Asymmetry," *Négotiations* 16, no. 2 (2011): 39-56, https://doi.org/10.3917/neg.016.0039.
[75] Mahmood Mamdani, *Neither Settler nor Native: The Making and Unmaking of Permanent Minorities* (Cambridge: Harvard University Press, 2020), https://www.hup.harvard.edu/catalog.php?isbn=9780674987326.

In conclusion, we therefore offer some key questions and areas for further enquiry that may help shape a DSD research agenda. A non-exhaustive list of questions could include:

*Conceptual and Operational*
- How does DSD differ or align across sectors, geographies, communities, and contexts?
- What can be learned from other (self-)governance practices in the further development and enforcement of DSD?
- What are the conditions and drivers that can enable a principled implementation of DSD?

*Processes*
- What design principles should inform the creation and implementation of "data assemblies" or other deliberative processes?
- What can be learned from recent deliberative democracy practices and/or innovations in collective bargaining processes to operationalize DSD?

*Policies*
- How are the components of a possible charter or statement of intent regarding DSD? And who should be involved in drafting these?
- How to deepen and operationalize the concept of "social license" across contexts and sectors?
- What can be learned from existing code-of-conducts in the development of a DSD code of conduct?

*People and Organizations*
- How to train "data stewards" who have a responsibility to define and comply with DSD conditions?
- What is the role of existing institutions and intermediaries (such as unions, community organizations and others) in representing vulnerable groups when DSD is negotiated or determined?
- How to ensure that new disintermediaries do not become new choke-points?

*Products and tools*
- What guidelines should be in place to steer transnational trustworthy data spaces that can provide legal certainty and accountability?
- How can products and tools be used to bridge gaps in digital literacy to empower DSD?

**Competing interests:** the authors declare none
**Data availability:** Not applicable to this paper
**Funding statement:** The GovLab received support from the Swiss Federal Government to hold some studios on migration and Digital Self determination; and I spend a month at the Rockefeller Foundation's Bellagio Center where the current draft was written